\documentclass[preprint,superscriptaddress,showpacs,preprintnumbers,amsmath,amssymb,aps, pre]{revtex4-1}

\usepackage[utf8]{inputenc}
\usepackage[english]{babel}
\usepackage{subfigure} 
\usepackage{graphicx}
\usepackage{dcolumn}
\usepackage{bm}

\begin{document}

\title{Chaotic dynamics and fractal structures in experiments with cold atoms}

\author{Alvar~Daza}
 \address{Nonlinear Dynamics, Chaos and Complex Systems Group, Departamento de  F\'isica, Universidad Rey Juan Carlos\\
Tulip\'an s/n, 28933 M\'ostoles, Madrid, Spain}

\author{Bertrand~Georgeot}
\address{Laboratoire de Physique Th\'eorique, IRSAMC, Universit\'e de Toulouse, CNRS, UPS, France}

\author{David~Gu\'ery-Odelin}
\address{Laboratoire Collisions, Agr\'egats, R\'eactivit\'e, IRSAMC, Universit\'e de Toulouse, CNRS, UPS, France}

\author{Alexandre~Wagemakers}
\address{Nonlinear Dynamics, Chaos and Complex Systems Group, Departamento de  F\'isica, Universidad Rey Juan Carlos\\
Tulip\'an s/n, 28933 M\'ostoles, Madrid, Spain}
 
\author{Miguel A.F.~Sanju\'an}
\address{Nonlinear Dynamics, Chaos and Complex Systems Group, Departamento de  F\'isica, Universidad Rey Juan Carlos\\
Tulip\'an s/n, 28933 M\'ostoles, Madrid, Spain}

\date{\today}

\begin{abstract}
We use tools from nonlinear dynamics to the detailed analysis of cold atom experiments. A powerful example is provided by the recent concept of basin entropy which allows to quantify the final state unpredictability that results from the complexity of the phase space geometry. We show here that this enables one to reliably infer the presence of fractal structures in phase space from direct measurements. We illustrate the method with numerical simulations in an experimental configuration made of two crossing laser guides that can be used as a matter wave splitter. 
\end{abstract}

%\pacs{05.45.-a, 05.45.Df}
\maketitle

\section{\label{sec:Introduction}Introduction}

Splitters for guided propagating matter waves were thoroughly investigated in the thermal regime \cite{cassettari_beam_2000,renn_laser-guided_1995,muller_magnetic_2001,houde_cold_2000,dumke_micro-optical_2002} in the past few years. More recently, the Bose-Einstein condensate regime was explored using optical waveguides \cite{gattobigio_exploring_2011,gattobigio_optically_2012}. Despite the quantum nature of these systems, some results could be understood using classical mechanics. For instance, a classical approximation was employed to unveil the chaotic dynamics underlying the experimental results in Ref.~\cite{gattobigio_optically_2012}. Positive Lyapunov exponents, a hallmark of chaos, were also estimated for experiments with ultracold atoms \cite{gattobigio_exploring_2011}. The recent development of new techniques of nonlinear dynamics allows to extract much more information from the same kind of experiments and, more importantly, to indicate how we can infer these signatures of chaos not only by numerical simulations but from direct measurements.

One of these powerful tools is the basin entropy, introduced in \cite{daza_basin_2016}. The framework of the basin entropy allows to quantify the unpredictability associated to the different outcomes in a dynamical system. Here we extend the basin entropy methodology to scattering problems. Using only the data that could be measured in real experiments such as the crossed laser beams setting of Ref.~\cite{gattobigio_optically_2012}, we can classify the complexity of the phase space for different parameters. Specifically, the basin entropy takes into account variations in the geometry of the phase space and the number of escapes. 
These factors contribute to the final state predictability of the system. We show that our approach provides a methodology for the reliable detection of fractal structures in phase space, given an experimental resolution. Using the same data set it is also possible to test the Wada property \cite{kennedy_basins_1991,daza_testing_2015}, a more restrictive property than fractality.

The experimental investigation of fractality is usually considered as a tour de force since data should be collected for a very large variation of the parameters (over a few orders of magnitude). The method presented here and based on the concept of basin entropy circumvents partially this difficulty \cite{daza_basin_2016} by defining an indicator at a given scale which detects faithfully fractality in most cases. We propose to use this concept to characterize the phase space fractal properties in a scattering problem. More specifically, we consider the configuration experimentally investigated in Refs. \cite{houde_cold_2000,dumke_micro-optical_2002,gattobigio_exploring_2011,gattobigio_optically_2012} and propose a new approach to such experiments that exploit a Monte Carlo sampling of the incoming wave packet parameters combined with a proper statistical analysis.

 The organization of the article is as follows. In Sec.~\ref{sec:Model}, we introduce the experimental setup made of two crossing guides and its modeling. Section \ref{basinentropy} is devoted to the adaptation of the concept of basin entropy to scattering situations. The application of this technique to the considered experimental cold atom system is detailed in Sec.~\ref{application}. In Sec.~\ref{sec:fractal}, we explain how this approach allows to experimentally characterize the fractal geometry of the phase space. Finally, a discussion about the application of these techniques to real experiments is presented.

\section{\label{sec:Model}The crossed beam configuration}

\begin{figure}
\begin{center}
\subfigure[~]{\includegraphics[width=7.5cm]{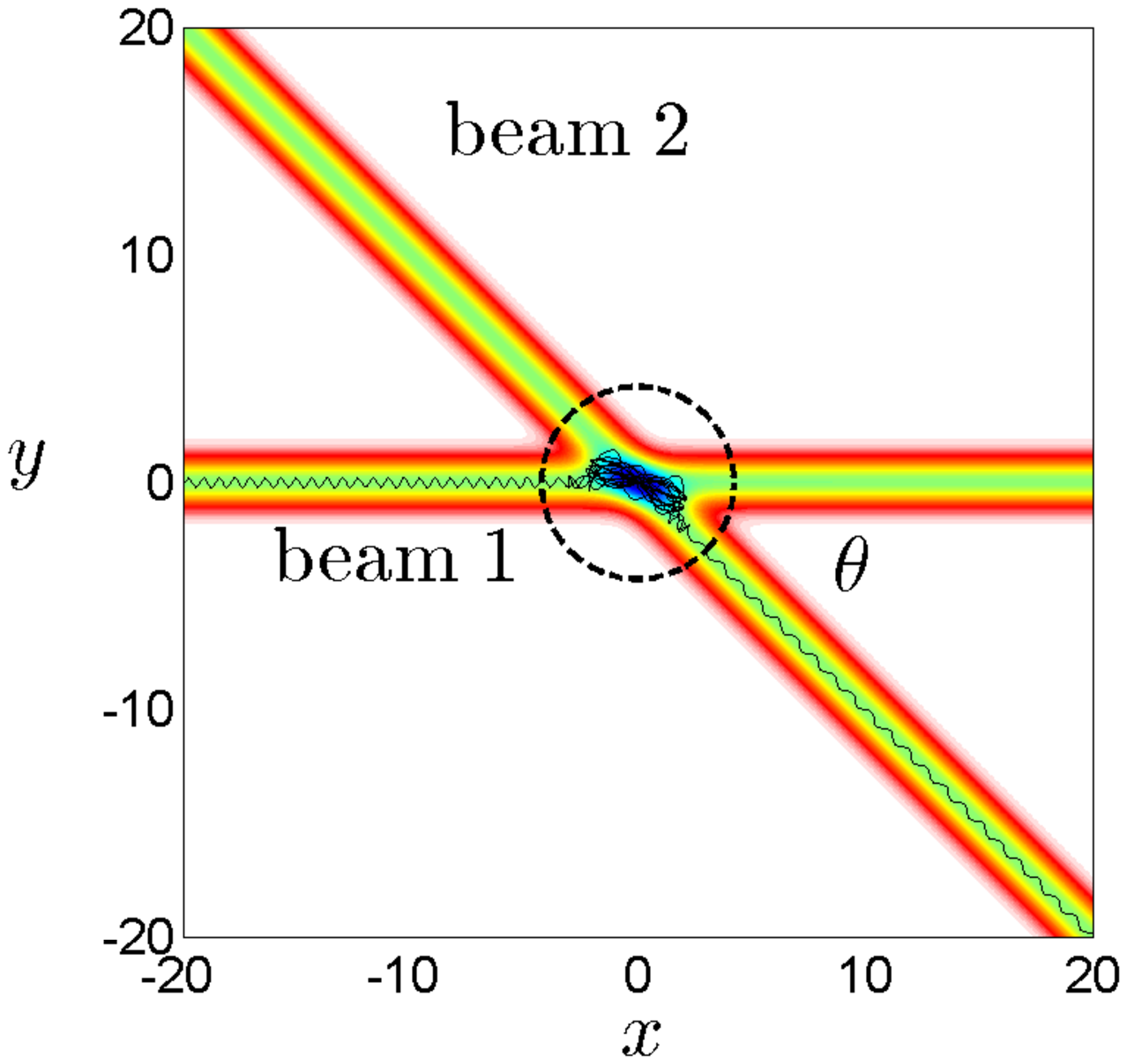}}
\subfigure[~]{\includegraphics[width=7.5cm]{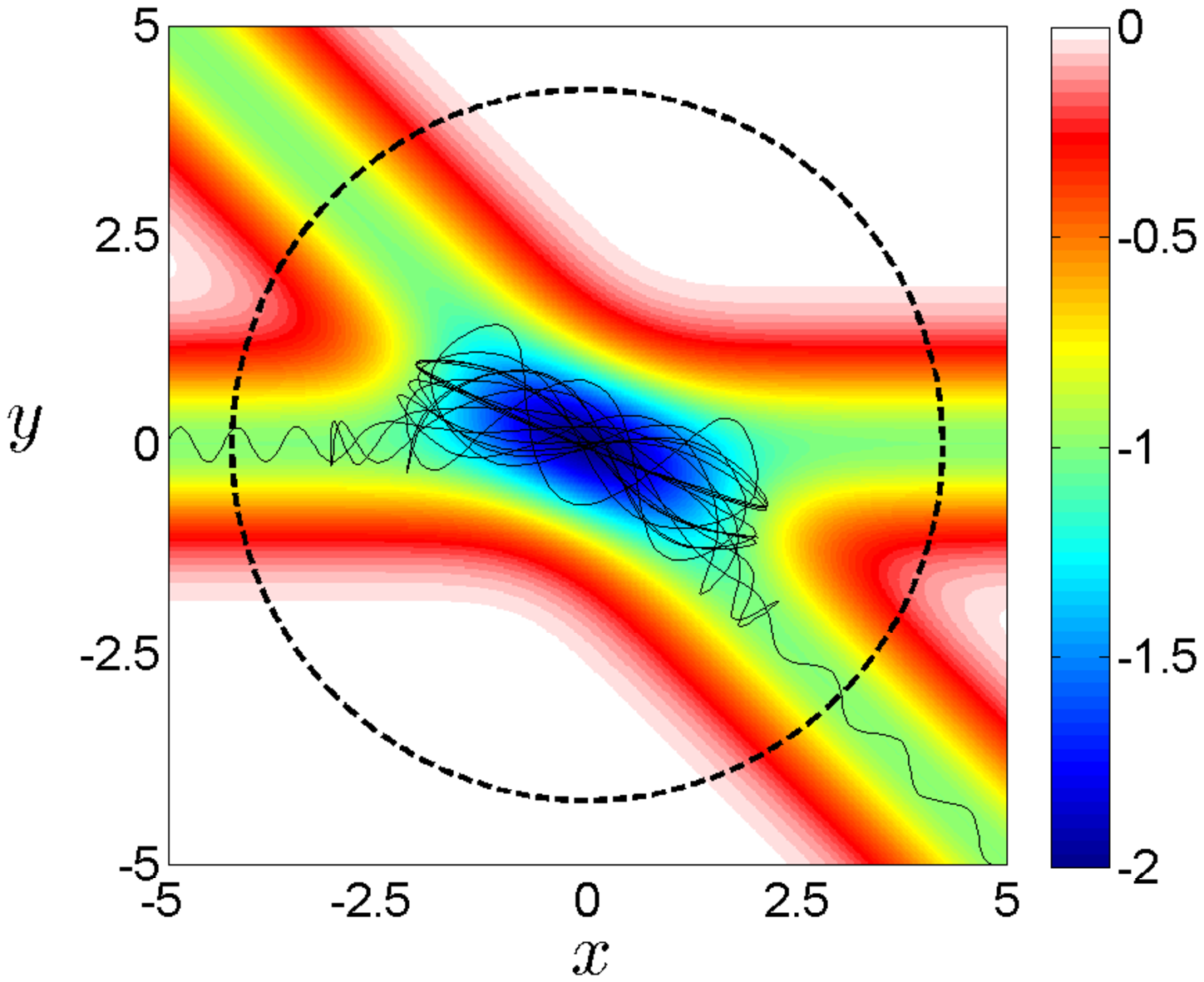}}
\end{center}
\caption{(Color online) \label{fig:Trajectory} \textbf{Example of a trajectory in the crossed beam configuration} ($x$ and $y$ are dimensionless).(a) An atom is shot from the beam 1 and its trajectory is depicted in black. The color code (grayness) accounts for the depth of the potential, being deeper for cold colors. (b) A zoom in the crossing region showing chaotic trajectories. The parameters of the Hamiltonian (\ref{eq:Hamiltonian}) are $\alpha_1=\alpha_2=\beta_1=\beta_2=1, \theta=45^{\circ}$. The dashed circle represents the scattering region $\sqrt{x^2+y^2}<3\sigma$, with $\sigma=\sqrt{\frac{2}{\beta_i}}$.}
\end{figure}

In this section we introduce the system studied along this work.  It corresponds to the motion of atoms into two crossed waveguides (see Fig.~\ref{fig:Trajectory}(a)). The scattering of an incoming atom laser outcoupled from a Bose-Einstein condensate in this specific configuration has been investigated in Ref.~\cite{gattobigio_optically_2012}. This type of system is quite generic in guided atom optics to generate beam splitters \cite{cassettari_beam_2000,renn_laser-guided_1995,muller_magnetic_2001,houde_cold_2000,dumke_micro-optical_2002,gattobigio_exploring_2011,gattobigio_optically_2012}. In practice, the waveguides can be produced by light using two out-of-resonance Gaussian laser beams crossing at an angle $\theta$ \cite{houde_cold_2000,dumke_micro-optical_2002,gattobigio_exploring_2011,gattobigio_optically_2012}. They generate a potential energy of the form
\begin{equation}
U(x,y,z)=-U_1\frac{w_{10}^2}{w_1^2(x)}e^{-2(y^2+z^2)/w_1^2(x)}-U_2\frac{w_{20}^2}{w_2^2(x^\prime)}e^{-2(y^{\prime\,2}+z^2)/w_2^2(x^\prime))}
\label{eq:pote}
\end{equation}
with $x^\prime=x\cos \theta-y\sin\theta$, $y^\prime=x\sin \theta + y\cos\theta$, $U_i=\eta P_i/w_{i0}^2$ and $w_i^2(x)=w_{i0}^2 (1+x^2/x_{Ri}^2)$ for $i=1,2$. Subindices 1 and 2 account for the two lasers, while $\theta$ is the angle between them. The parameter $w_{i0}$ refers to the waist of the Gaussian dipole laser $i$, $\lambda_i$ to its wavelength, and $x_{Ri}=\pi w_{i0}^2/\lambda_i$ to its Rayleigh length. The $\eta$ parameter has a value that depends both on the atom and on the wavelength of the dipole laser ($\eta=1.3 \times 10^{-36}$ JW$^{-1}$m$^2$ for rubidium-87 with $\lambda=1064$ nm \cite{grimm_optical_2000}).

For the sake of simplicity, we shall use a two-dimensional model that captures the main features of the experimental system \cite{gattobigio_optically_2012,torrontegui_cold-atom_2010}. For this purpose, we assume that the propagation is performed on a small distance with respect to the Rayleigh length so that $w_i^2(x)\simeq w_{i0}^2$. Only one transverse direction $y$ is considered and $x$ is the direction of propagation. We follow a dimensionless procedure by introducing the length scale $\ell$ and the time scale $\tau$ in Eq.~(\ref{eq:pote}). We define $\alpha_i=mU_i\tau^2/\ell^2$ and $\beta_i=2\ell^2/w_{i0}^2$, and redefine $x/\ell \rightarrow x$ (resp. $y/\ell \rightarrow y$) yielding the following Hamiltonian with dimensionless variables:

\begin{equation}
H=\frac{1}{2} \left( \dot{x}^2+\dot{y}^2\right)  -\alpha_1 e^{-\beta_1 y^2}-\alpha_2 e^{-\beta_2 (x\sin\theta+y\cos\theta)^2}.
\label{eq:Hamiltonian}
\end{equation}

Now, the features of each laser are condensed into two characteristic parameters: $\alpha$, related to the depth of the potential and $\beta$, related to the laser waist. Along this work, we will use $\alpha_i=\beta_i=1$, which means $w_{i0}=w_{0}$, $U_i=U_0$ and therefore $\ell=w_0\sqrt{2}$ and $\tau=w_0 (2/mU_0)^{1/2}$.

Figures~\ref{fig:Trajectory}(a)-(b) show an example of a classical trajectory of this Hamiltonian. The coupling of the longitudinal and the transverse degrees of freedom that occurs at the crossing region is responsible for the complex dynamics.
The classical description could surprisingly account for the experimental results, as shown in \cite{gattobigio_optically_2012}. The physical reasons are twofold: 1) the typical scale of variation of the potential is large compared to the de Broglie wavelength associated with the incoming velocity, and 2) interference effects were marginal because of the relatively short time that the wave packet spends in the scattering region, and the 3D dynamics limiting the overlap of the packet with itself. In the following, we shall investigate the fractal properties of this system using a paving of the classical phase space. Our results remain pertinent for the experiments once the phase space cells considered for the statistical analysis are significantly larger than $\hbar$.

\begin{figure}
\begin{center}
\includegraphics[width=15cm]{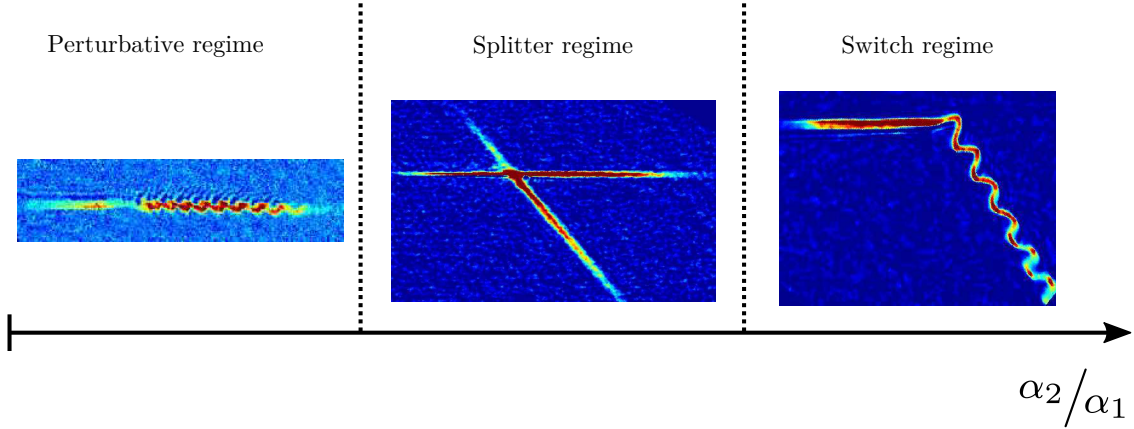}
\end{center}
\caption{(Color online)\label{fig:Exp} \textbf{Qualitative presentation of the different experimental regimes.} In Ref.~\cite{gattobigio_optically_2012}, where atom lasers propagate in a X-shape configuration, three regimes were observed depending on the power ratio $\alpha_2/\alpha_1$ (see experimental pictures). The position of the boundaries between these different regimes also depends on the initial horizontal speed $v_{x0}$. Here we keep $\alpha_2/\alpha_1=1$ and modify $v_{x0}$, which is also feasible in real experiments.}
\end{figure}

In that work \cite{gattobigio_optically_2012}, the parameters governing the potential in Eq.~\ref{eq:Hamiltonian}, i.e., the parameters of the waveguides $\alpha_1,\alpha_2,\beta_1,\beta_2$, were changed in order to produce different kind of dynamics. In particular, for low values of the ratio of intensities $\alpha_2/\alpha_1$ the waves were only slightly perturbed, for similar intensities of both lasers $\alpha_2/\alpha_1\approx1$ the splitter regime was found, and finally, for large values of $\alpha_2/\alpha_1$ the switch regime was dominant (see Fig.~\ref{fig:Exp}). In the present paper, we adopt a different perspective by keeping the potential unchanged ($\alpha_1=\alpha_2=\beta_1=\beta_2=1$) and varying the initial horizontal speed $v_{x0}$, which is more in the spirit of the scattering problems. This is feasible in experiments by setting different gradients to accelerate the atoms \cite{vermersch,damon}. The initial horizontal speed $v_{x0}>0$ can be considered as a parameter of the system: the dynamics depends strongly on this value. Namely, when particles are shot with a low speed $v_{x0}$, trajectories have more time to explore the scattering region and to display chaotic dynamics. For high speed $v_{x0}$, particles are barely affected by the potential.

\begin{figure}
\begin{center}
\subfigure[~]{\includegraphics[width=7cm]{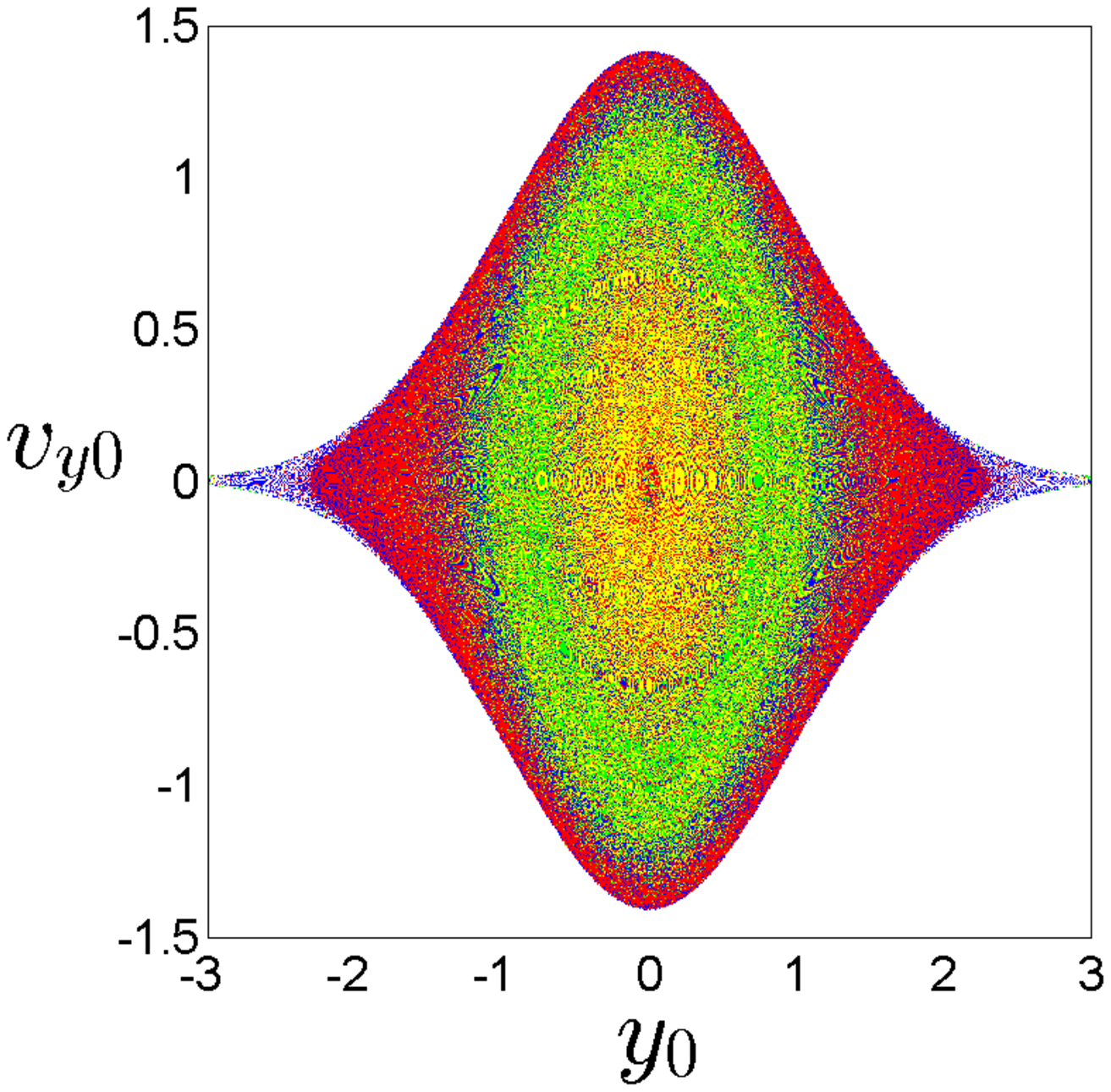}}
\subfigure[~]{\includegraphics[width=7cm]{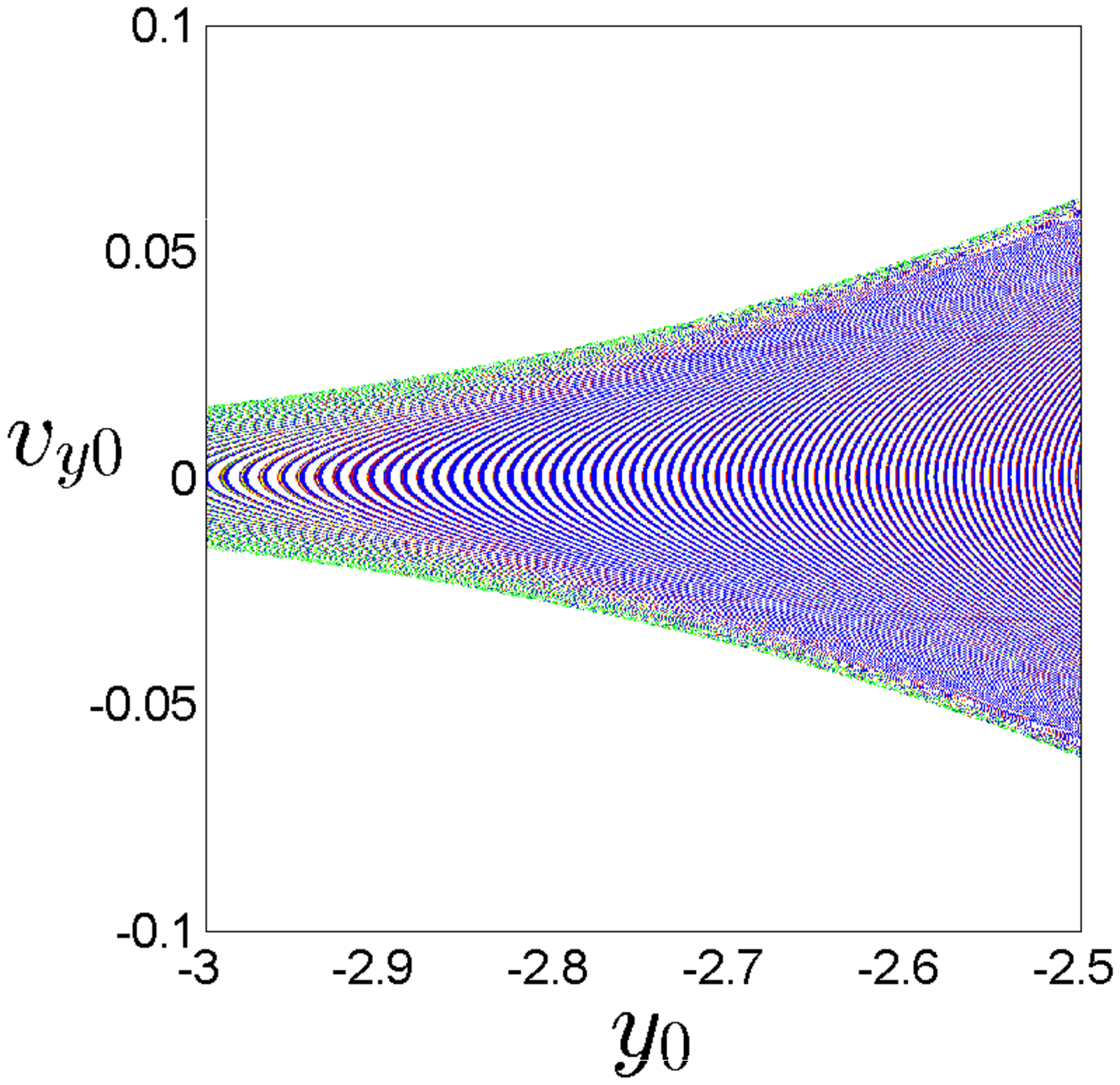}}
\subfigure[~]{\includegraphics[width=7cm]{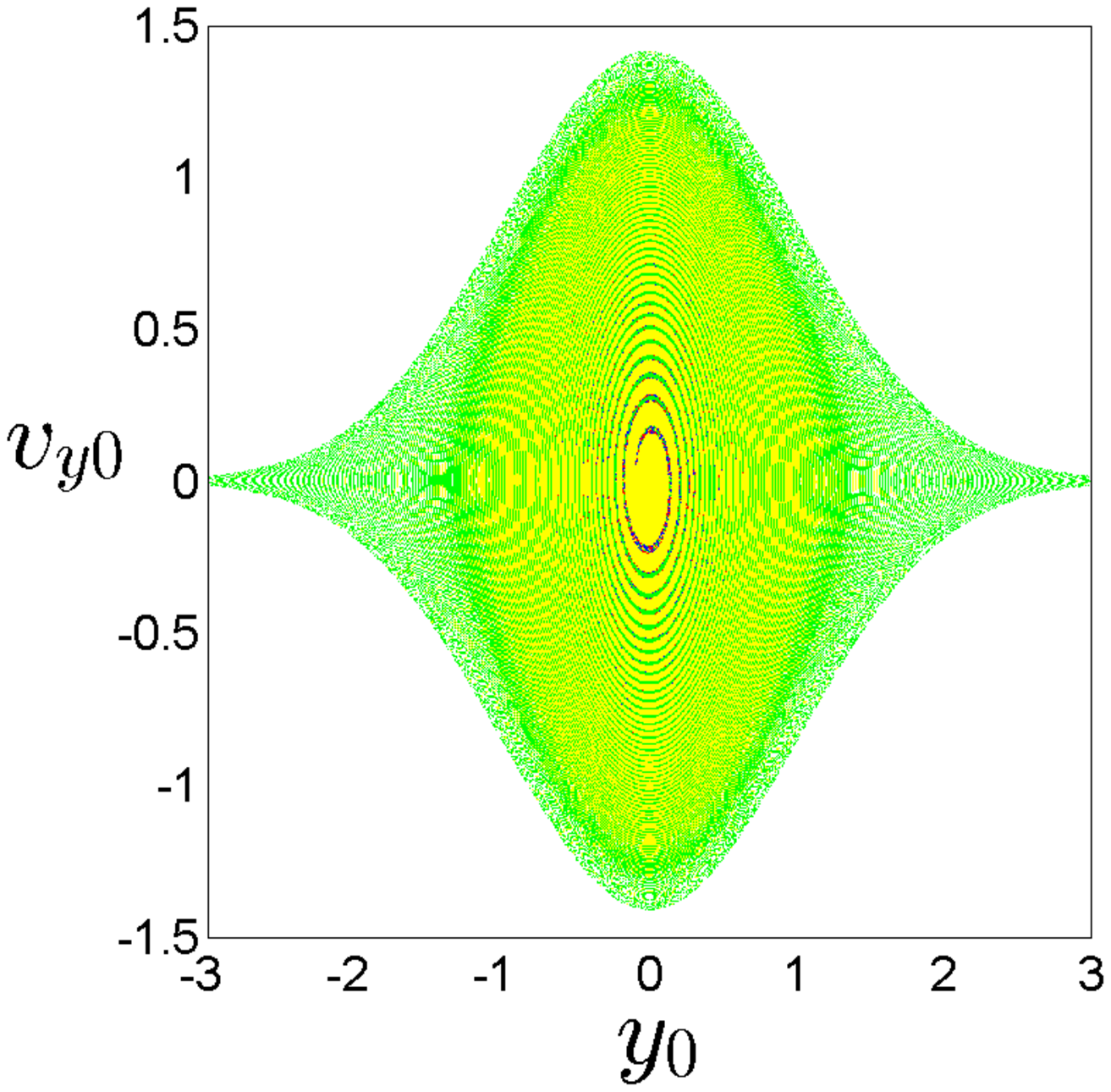}}
\subfigure[~]{\includegraphics[width=7cm]{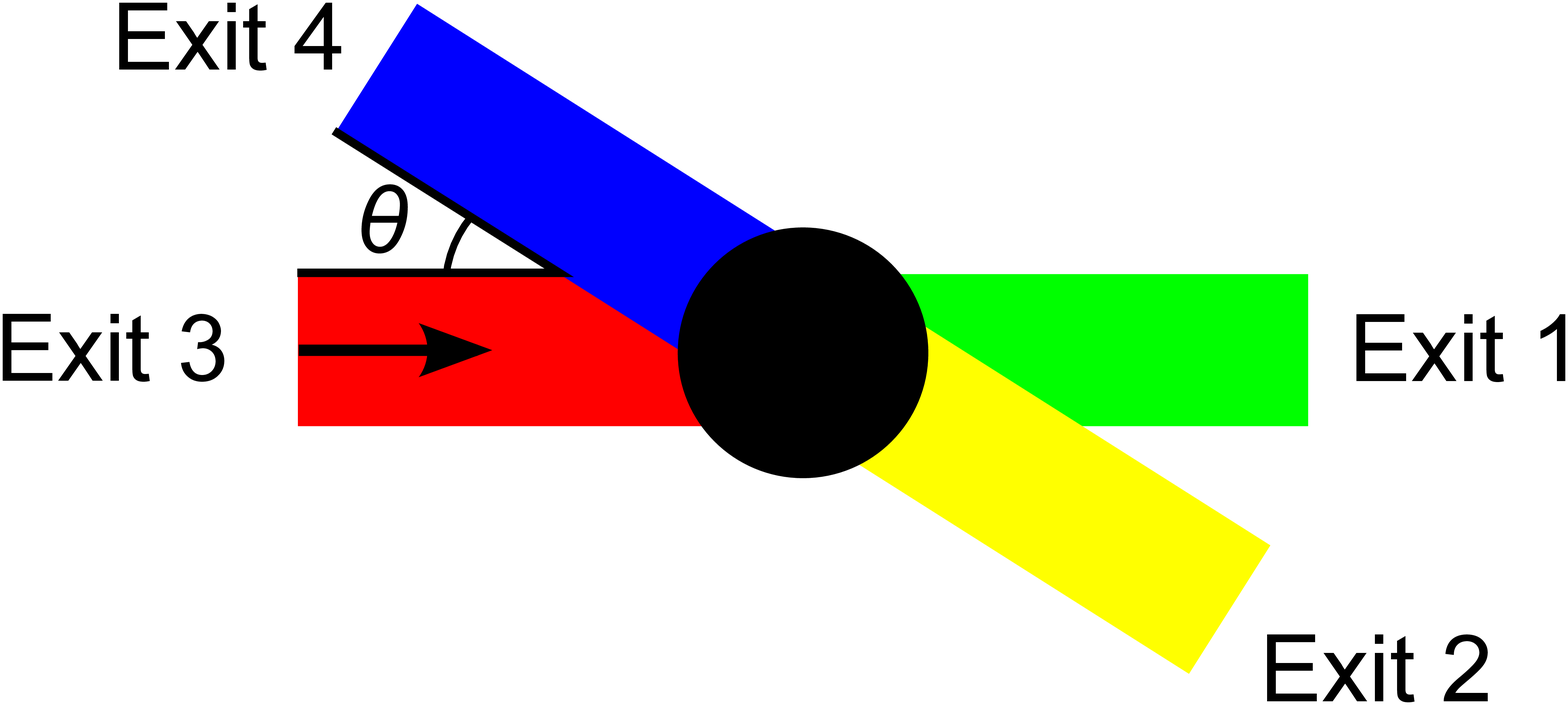}}
\end{center}
\caption{(Color online) \label{fig:Basin} \textbf{Escape basins}  ($y_0$ and $v_{y0}$ are dimensionless). The parameters for these basins are $\theta=45^{\circ},x_0=-500,\alpha_1=\alpha_2=\beta_1=\beta_2=1$. (a) Escape basin for a low shooting speed $v_{x0}=0.1$. (b) Zoom in the basin depicted in (a). (c) Escape basin for a high shooting speed $v_{x0}=1$. (d) Color code used to label the basins.}
\end{figure}

Therefore, by considering $v_{x0}$ and $\theta$ as parameters and assuring sufficiently long shooting distances $x_0$ from the crossing region, we can analyze the dynamics in terms of $(y_0,v_{y0})$. The set of initial conditions $(y_0,v_{y0})$ that yields an escape through a given exit is referred to as an escape basin \cite{aguirre_wada_2001}. Given the Gaussian profile of the potentials, we define unbounded trajectories as those going further than $3 \sigma_i$ of each laser beam $i=1,2$, with $\sigma_i=\sqrt{\frac{2}{\beta_i}}$. An example of such scattering region is delimited in Fig.~\ref{fig:Trajectory} by dashed lines. Graphical representations of escape basins are provided in Fig.~\ref{fig:Basin}-(a)-(c), where each color represents an exit according to the color code of Fig.~\ref{fig:Basin}-(d).  White pixels are for such unbounded trajectories that provide atom losses and also for what we call sticky trajectories, i.e., that they spend more than $2\cdot10^6$ time steps without escaping. These two kinds of trajectories will not be considered for the calculations of the basin entropy due to their negligeable influence. Their corresponding basin is however interwoven with the other basins, as shown in Fig.~\ref{fig:Basin}-(b), but it is only important for extremely low values of $v_{y0}$ and large initial transverse positions $y_0$. In the following, we will restrict our study to the parameter ranges $v_{y0}\in[-1.5,1.5]$,  $y_0\in[-1.5,1.5]$. 

The presence of fractal structures is noticeable for low speed basins (see Fig.~\ref{fig:Basin}-(a)), but harder to appreciate in the case of high speed, e.g.,  Fig.~\ref{fig:Basin}-(c). Quantifying the different \textit{degrees of fractality} can be done with the help of basin entropy.

\section{The concept of basin entropy for scattering experiments}
\label{basinentropy}

The idea of basin entropy \cite{daza_basin_2016} was introduced in order to answer a simple question: how is it possible to affirm that one basin is more unpredictable than another? To this aim, a new quantitative measure was defined. To calculate the value of the basin entropy, we proceed in several steps.  First, the phase space must be divided into smaller parts, that we call boxes. Each box contains in principle infinitely many different initial conditions, leading to a finite number of final destinations. In every box $i$, the proportion of initial conditions leading to each final destination $j$ defines the probability $p_{i,j}$ of a trajectory inside that box ending in that particular final destination. By means of the Gibbs entropy we can define the entropy of a box as $S_i=-\sum\limits_{j=1}^{N_A} p_{i,j}  \ln p_{i,j}$, where $N_A$ is the number of possible final destinations of the system, and for our scattering problem the number of exits. The two guides give rise to four possible exits, that is $N_A=4$ in our case, so here the entropy of every box $S_i$ ranges between 0, if all the initial conditions inside a box lead to the same exit, to $\ln N_A$ for an equiprobable distribution of the initial conditions inside the box. Next, we add the entropies of all the boxes considered $S=\sum\limits_{i=1}^{N}S_i$. Finally, in order to get a quantity ranging from 0 to $\ln N_A$, we divide that quantity $S$ by the total number of boxes $N$, obtaining the basin entropy $S_b={S}/{N}$.

\begin{figure}
\begin{center}
{\includegraphics[width=7.5cm]{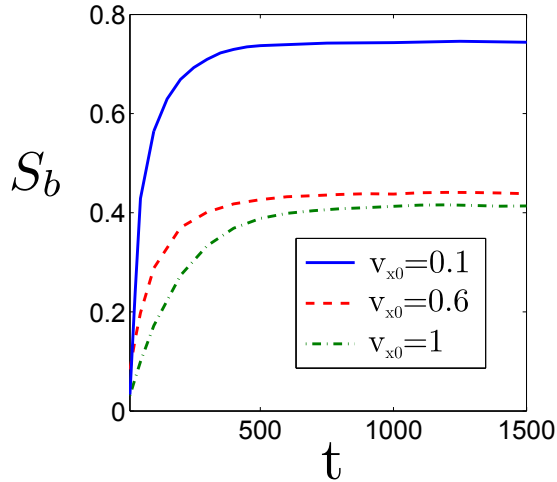}}
\end{center}
\caption{(Color online) \label{fig:diff_x0} \textbf{Basin entropy for different launching distances.} The angle of the beams is $\theta=45^{\circ}$, $v_{y0}\in[-1.5,1.5]$ and  $y_0\in[-1.5,1.5]$. Other parameters are $\alpha_1=\alpha_2=\beta_1=\beta_2=1$, as used throughout the paper. For each horizontal speed $v_{x0}$, we compute the basin entropy $S_b$ for different launching distances $x_0$, so that the atoms take a time $t$ to reach the scattering center in $x=0$. In our regime of parameters, the basin entropy saturates for a time $t\approx500$ regardless of the initial horizontal speed $v_{x0}$.}
\end{figure}

The basin entropy can be computed numerically by integrating the equations of motion and obtaining the escape basins. This quantity can also be inferred from experimental data, as we will show in the next section. But before that, we must discuss some technical details concerning the basin entropy calculation in scattering problems. First, when we compute numerically the basin entropy we usually work with flat distributions, in the sense that initial conditions are uniformly distributed in the four dimensional space $(x_0,v_{x0},y_0,v_{y0})$. When particles advance through the horizontal waveguide, before arriving to the scattering region, these distributions evolve in time. Indeed, the transverse Hamiltonian is close to integrability and possesses invariant curves which do not correspond to the original distribution. Trajectories follow these invariant lines towards a quasi stationary regime where they are uniformly distributed on these curves. This happens for sufficiently long times, i.e., for sufficiently long launching distances. If the particles do not have enough time to evolve to these asymptotic distributions, the values of the basin entropy can be largely modified, but for long enough times the basin entropy remains nearly as a constant for every initial shooting speed $v_{x0}$, as shown in Fig.~\ref{fig:diff_x0}.

Another different issue is the number of trajectories per box. In order to get a reliable value of the basin entropy, it is necessary to have reliable values of the probabilities of each four-dimensional box. However, in numerical simulations we must reach a compromise between computational cost and statistical convergence. Results remain unaltered for values of the trajectories per box larger than $5^4$ (a hypercube of side 5 in the four-dimensional space sampled), which will be the standard value used along this work.

Finally, another fundamental parameter for the basin entropy calculation is the number of boxes $N$. The larger the number of boxes $N$, the more precise will be the description. However, both in experiments and in numerical simulations we must take into account practical considerations concerning the number of experiments and the computational effort. In this respect, it has already been shown \cite{daza_basin_2016} that a Monte Carlo procedure for the choice of the boxes in phase space leads to accurate values of the basin entropy minimizing the efforts. 

\section{Basin entropy from experimental cold atom data}
\label{application} 

\begin{figure}
\begin{center}
\includegraphics[width=7.5cm]{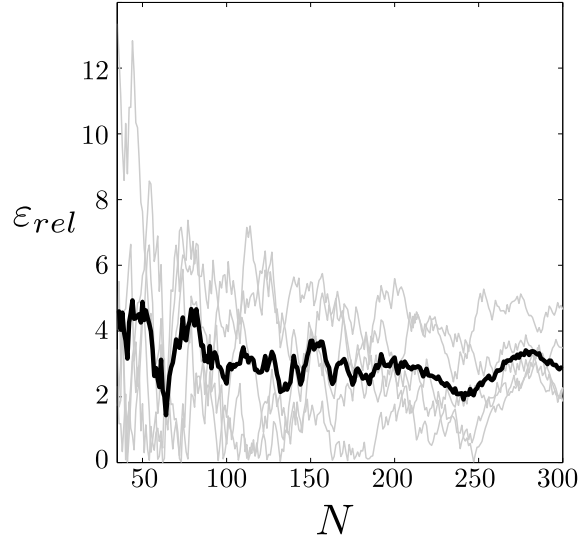}
\end{center}
\caption{\label{fig:Error} \textbf{Convergence with the number of boxes used in the Monte Carlo scheme.} The relative error of the basin entropy $\varepsilon_{rel}=\frac{\vert S_{b}-S_b(RS)\vert}{S_b} \times 100$  as the number of sampled boxes $N$ is represented.  $S_b$ is taken as a reference value of the basin entropy computed for $10000$ boxes, and $S_b(RS)$ is the basin entropy for the random sampling procedure. We can see that for a realistic number of experimental runs (50-100) the relative error is below $10\%$. In gray, five different runs made for $\theta=45^{\circ}$ and $v_{x0}\in[0.09,0.11]$,  $x_0\in[-250,-200]$, $v_{y0}\in[-1.5,1.5]$,  $y_0\in[-1.5,1.5]$. In black, the mean of the five runs in gray.}
\end{figure}

The procedure to calculate the basin entropy and the scattering experiments with cold atoms share some important similarities. In both cases we consider ensembles of trajectories instead of single trajectories. In the experiments we have clouds of atoms with different values of position and velocity, and for the basin entropy calculation we must compute many trajectories with different initial conditions inside every box. Scattering experiments essentially study the output of the trajectories in order to gain knowledge about the system, just as the basin entropy does. We propose to use as the equivalent of boxes in the basin entropy scheme, wave packets of atoms which are launched towards the scattering region. Indeed, these wave packets correspond to a group of atoms distributed around a mean value of the velocity and the position following a Gaussian distribution. The experimental measurement through absorption pictures provides access to the population of different branches, and thus to the probabilities inside every box. 

As we have described in the previous section, we must pay attention to some technical details concerning the basin entropy computation in scattering problems. In experiments, the problem of the stationarity of the distributions before its arrival to the scattering region can be solved by varying the launching distance appropriately. The number of trajectories per box is not a limit, since this number is related to the number of atoms in a wave packet, which in real experiments is in the thousands. In fact, it can be further increased by repeating the experiment for a wave packet with same initial mean values.

Remarkably, the Monte Carlo sampling of phase space can be done experimentally by selecting different sets of initial conditions with different mean velocity $v_{y0}$ and mean position $y_0$. In practice, small clouds of atoms shall be successively delivered from a trap that accommodates a reservoir of atoms such as a Bose-Einstein condensate placed upstream. The transverse position for outcoupling the atoms can be tuned by modifying with optical means the reservoir trap geometry, while the mean transverse velocity can be transferred to the packet of atoms by applying a well-calibrated transverse magnetic gradient pulse. The successive repetition of such outcoupling procedures until the reservoir is empty permits to reduce drastically the number of experimental runs. As shown in Fig.~\ref{fig:Error}, for a realistic number of experimental runs ($N\sim50$) the relative error in the basin entropy computation is below $10\%$.

Another important point is the size of the boxes used in the basin entropy computation, that is, the minimal resolution that can be reached in this experimental procedure. This corresponds to the size of the wave packet relative to the size of the range of phase space that we want to explore. To access the best achievable resolution, we focus on the velocity space since a similar argument will apply to the position space. Typically, the range of variation of the velocity is of the order of $\Delta v = \sqrt{U_0/m}$, associated to the depth $U_0$ of the guide. For a quantum packet in the transverse ground state, the velocity dispersion of the wave packet $\delta v_0=\omega_0 a_0$ where $a_0=(\hbar/m\omega_0)^{1/2}$ is the oscillator length and $\omega_0=(4U_0/mw_y^2)^{1/2}$ is the transverse angular frequency obtained by expansion of (\ref{eq:pote}). The realization of a monomode atom laser has proved the experimental feasibility of the production of such packets \cite{guerin,EPL2008,PRA2009,arlt,canberra,robins}. Using the parameters from Ref. \cite{gattobigio_optically_2012}, we find $\Delta v/\delta v_0 = (U_0mw_y^2/4\hbar^2)^{1/4}\simeq 65$ ($w_y=100$ $\mu$m and $U_0/k_B=10$ $\mu$K). In practice, a linear resolution of several tens can therefore be obtained up to a maximum of one hundred. This means that the experimental escape basins would have a resolution between $10\times10$ and $100\times100$.

In short, to compute the basin entropy $S_b$ in the crossed beam configuration, one should perform a sufficient number of experiments. Each of these experiments consists in sending a wave packet with some mean transversal velocity and position. The experiments must be carried out for sufficiently long launching distances to assure the stationarity of the distributions. Then, the population escaping through each channel should be measured by absorption images, for instance. Each experimental run provides a value of the basin entropy in a box $S_{i}$. With an appropriate sampling of the region of phase space considered, the total basin entropy can be computed by adding the basin entropy associated to each run.

\section{\label{sec:fractal}Chaos and fractal structures in escape basins}

\begin{figure}
\begin{center}
\subfigure[~]{\includegraphics[width=5cm]{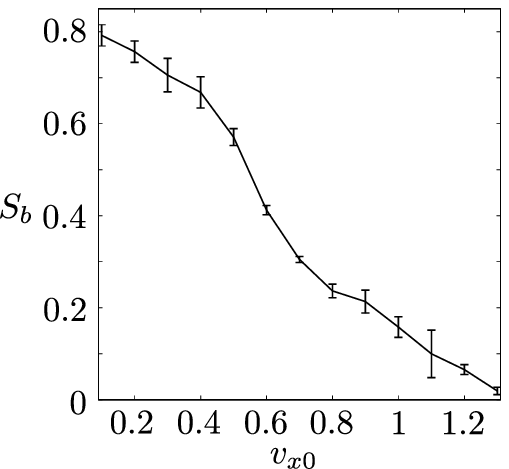}}
\subfigure[~]{\includegraphics[width=5cm]{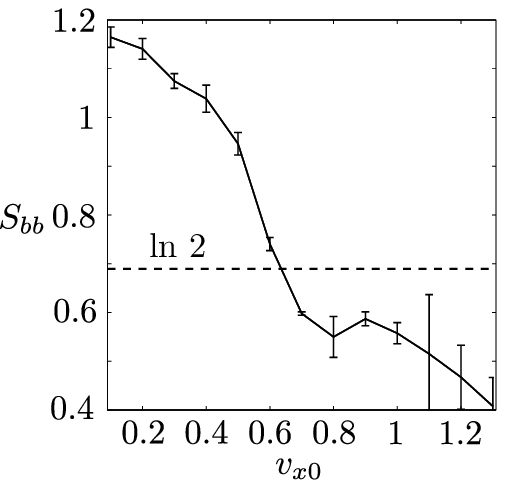}}
\subfigure[~]{\includegraphics[width=5cm]{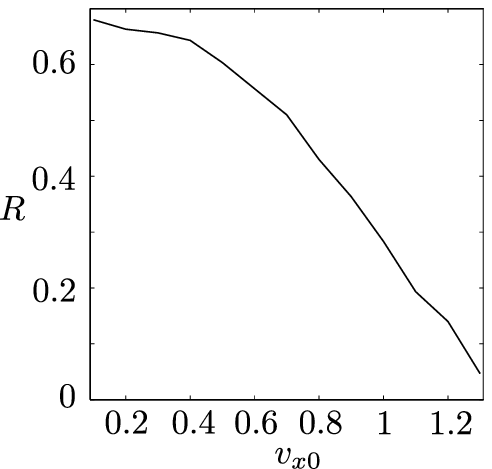}}
\end{center}
\caption{\label{fig:diff_vx0} \textbf{Random sampling computation using different values of the launching speed.} The volume of the phase space investigated is $v_{y0}\in[-1.5,1.5]$,  $y_0\in[-1.5,1.5]$, $x_0\in[-760,-750]$. For each mean value of $\left\langle v_{x0}\right\rangle$ we consider that $v_{x0}\in[\left\langle v_{x0}\right\rangle - \left\langle v_{x0}\right\rangle/10,\left\langle v_{x0}\right\rangle+\left\langle v_{x0}\right\rangle/10]$. The basin entropy $S_b$ is computed using 100 boxes (experimental runs) for each represented point, and this procedure is repeated three times so that we get the error bars displayed in the figures. (a) As we increase the horizontal speed $v_{x0}$ the basin entropy decreases. (b) The boundary basin entropy $S_{bb}$ is above the $\ln 2$ threshold (dashed line) for low speeds $v_{x0}$, and is below for high speeds. (c) The fraction of boxes, $R$, lying on the boundary decreases as we increase $v_{x0}$.}
\end{figure}

\begin{figure}
\begin{center}
\subfigure[~]{\includegraphics[width=7cm]{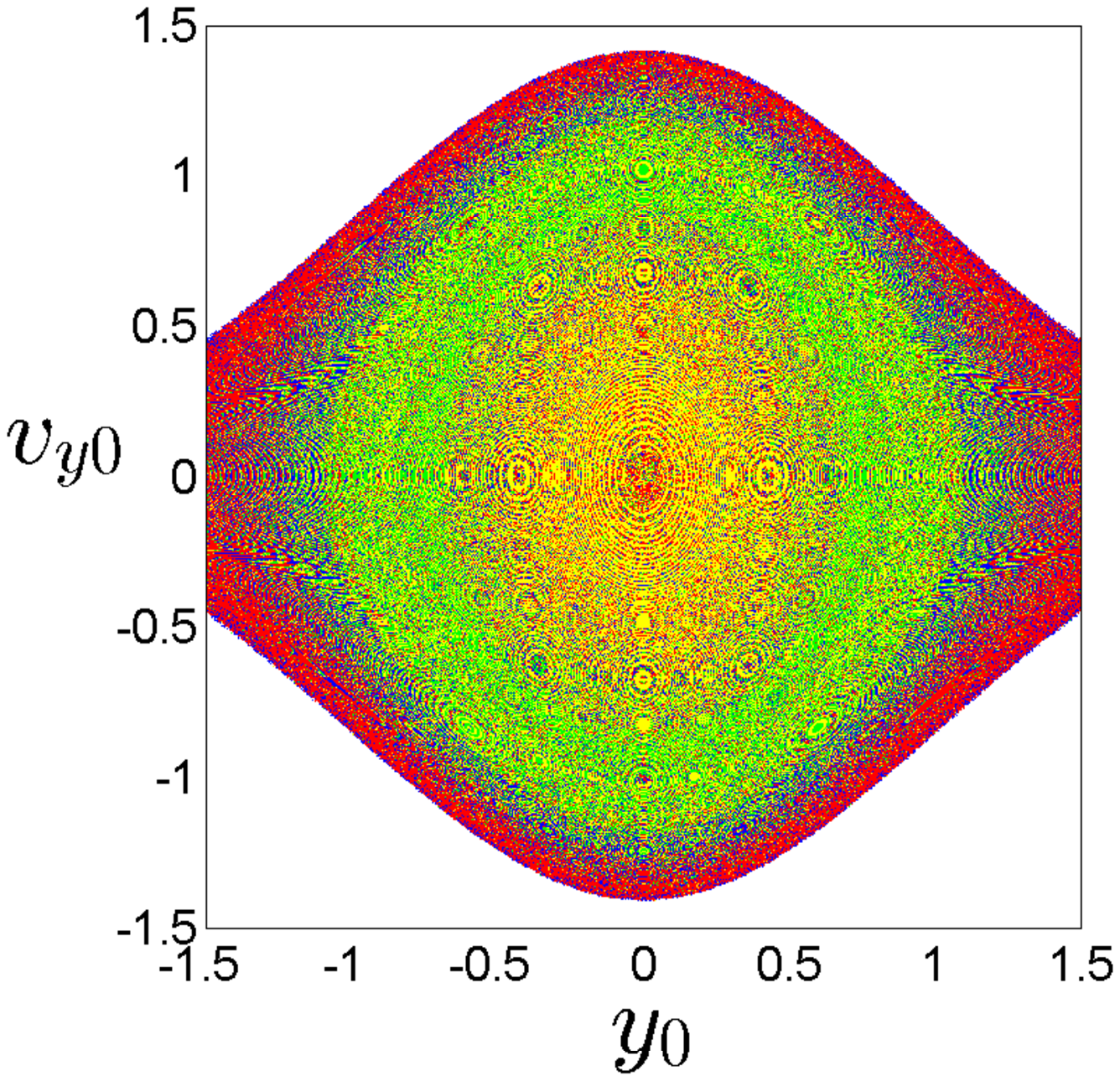}}
\subfigure[~]{\includegraphics[width=7cm]{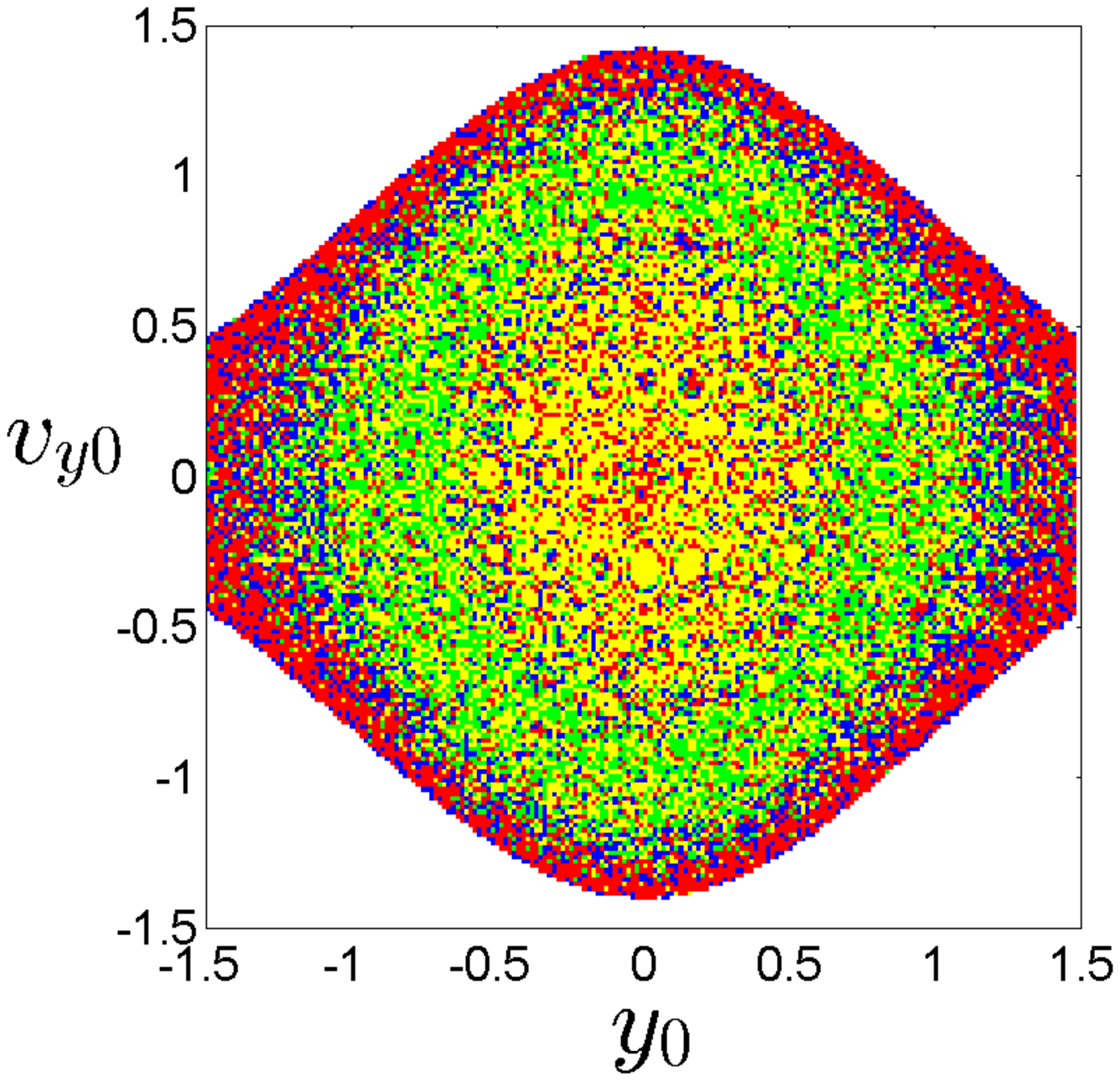}}
\subfigure[~]{\includegraphics[width=7cm]{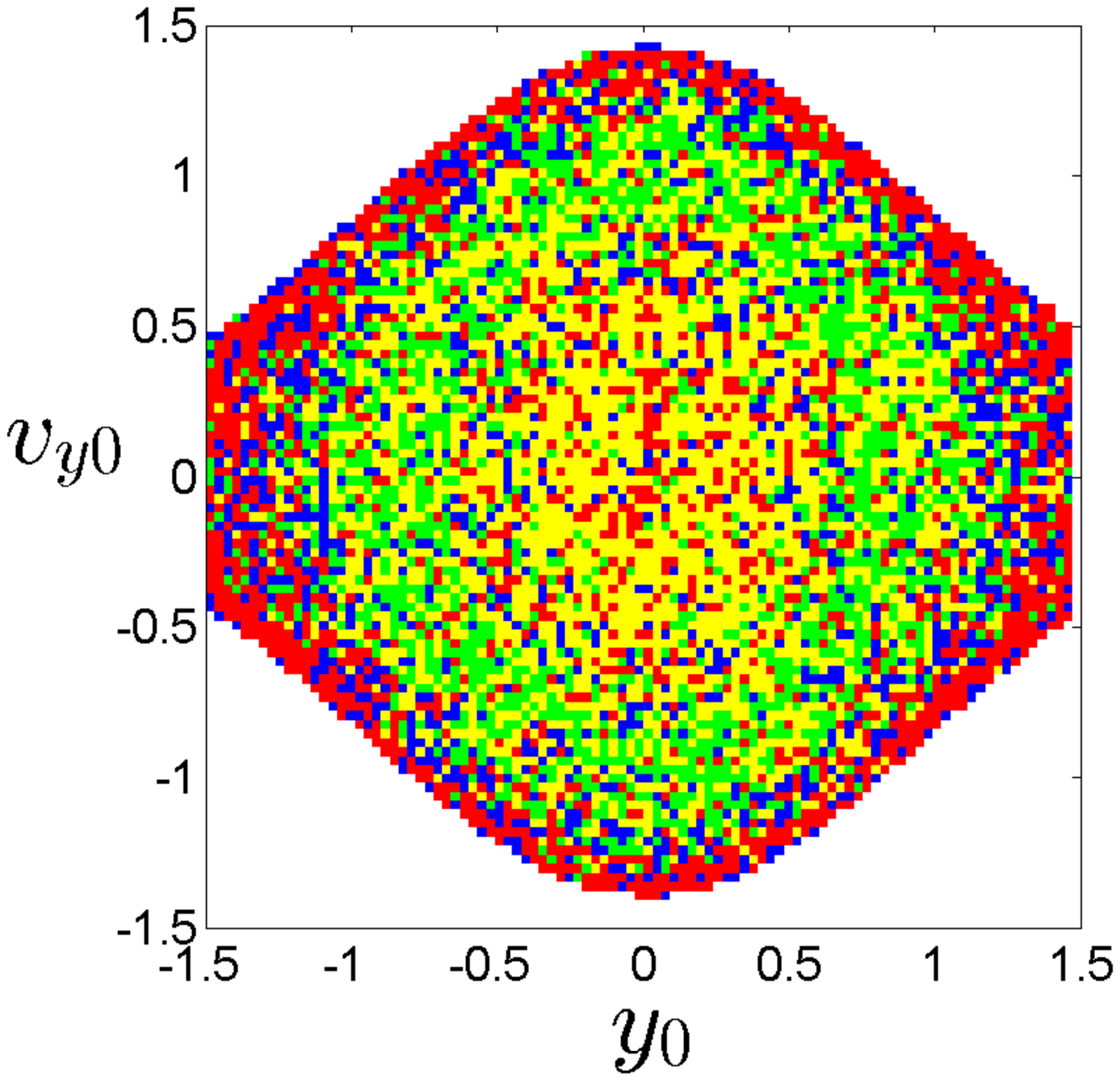}}
\subfigure[~]{\includegraphics[width=7cm]{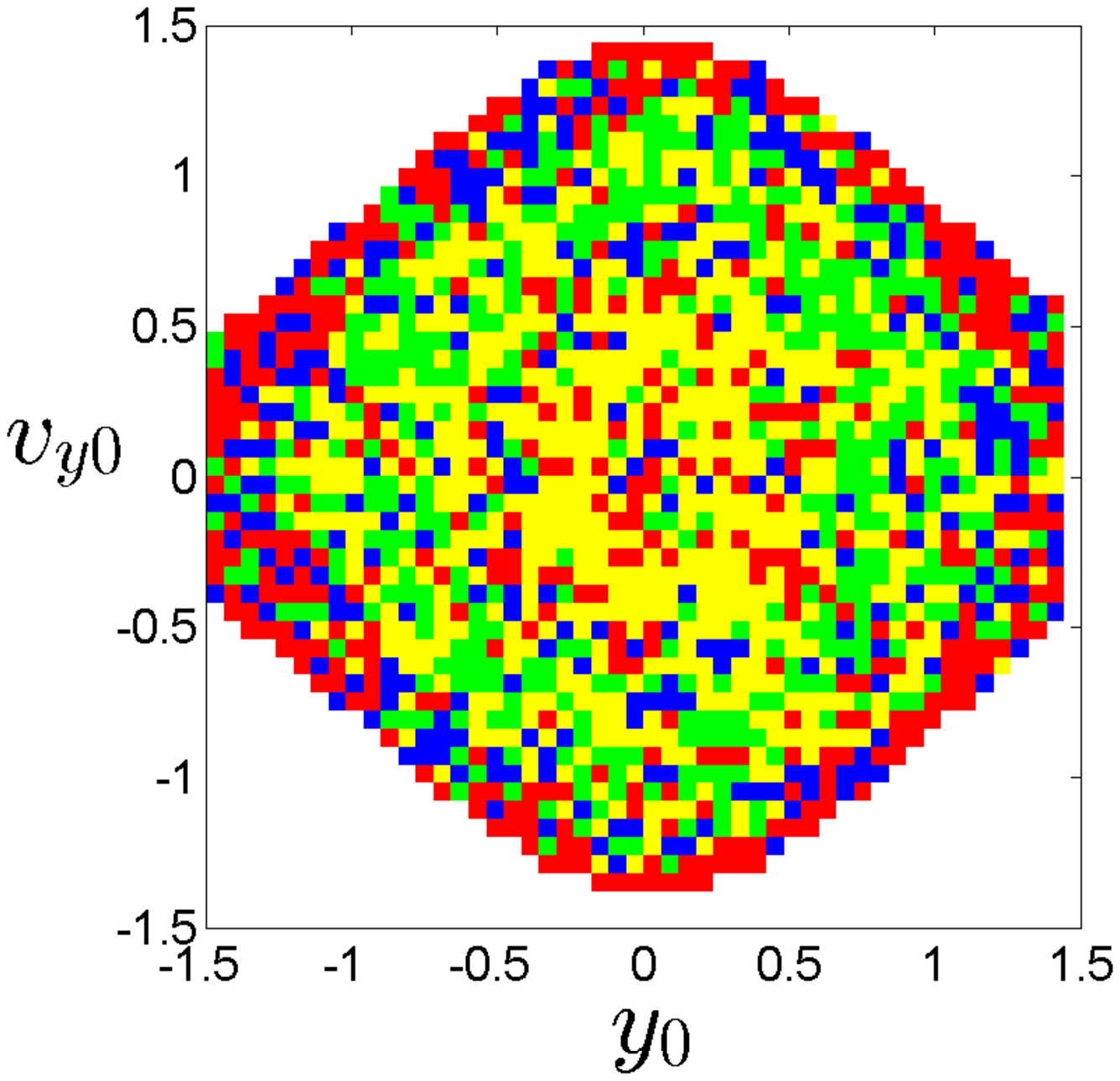}}
\end{center}
\caption{(Color online) \label{fig:resolution} \textbf{Escape basins for different resolutions.} We depict the escape basins in the $(y_0,v_{y0})$ subspace for $\theta=45^{\circ},x_0=-250,v_{x0}=0.1$. The resolutions are (a) $1000\times1000$, (b) $200\times200$, (c) $100\times100$ and (d) $50\times50$. The experimental resolution is probably close to (d), but still fractal structures are present.
}
\end{figure}

In this section, we investigate transient chaos and fractal structures appearing for low values of the horizontal velocity $v_{x0}$. A low speed implies that particles spend more time in the scattering region, i.e., the crossing region of the two beams. Therefore, the exponential divergence of trajectories induced by the complicated shape of the potential at the crossing makes the system difficult to predict. The basin entropy captures this strong unpredictability due to the \textit{highly fractalized} phase space. We have used a Monte Carlo procedure to sample the basin entropy $S_b$ for different shooting speeds $v_{x0}$ (see Fig.~\ref{fig:diff_vx0}-(a)). We can see that the basin entropy is higher for lower speeds, providing  us a quantitative basis to our intuition: it is more difficult to predict the final destination of particles with low speed $v_{x0}$.

Using these data, we can detect the presence of fractal structures in phase space. To this aim, we must compute the \textit{boundary basin entropy}, $S_{bb}$. This quantity is defined as $S_{bb}=S/N_b$ \cite{daza_basin_2016}, where $S$ comes again from $S=\sum\limits_{i=1}^{N}S_i$ and $N_b$ is the number of boxes containing more than one color. In other words, we are repeating the same procedure used to compute the basin entropy, but now we are normalizing only by the boxes lying on the boundary.

The quantity $S_{bb}$ therefore measures the complexity of the basin boundaries. Moreover, there is a threshold value of $S_{bb}$ that separates basins with smooth boundaries from those with fractal boundaries.
Indeed, suppose that our basins were separated by smooth boundaries. In that case, most of the boxes would contain boundaries separating only two basins, though there might be a finite number of boxes containing more than two basins. If we take a sufficiently large number of boxes, the contribution of these boxes separating more than two basins will be negligible. If the basins are smooth, the scenario leading to the largest boundary basin entropy, $S_{bb}=\ln 2$, corresponds to a situation for which all boundary boxes have equal proportions of the two basins.

Therefore, if the computed value of the boundary basin entropy obeys the inequality $S_{bb}>\ln 2$, then the boundaries must be fractal. This is referred to as the $\ln 2$ criterion \cite{daza_basin_2016}. 

The $\ln 2$ criterion is a sufficient but not necessary condition for fractality: some fractal basins do not pass this criterion, for instance those having only two outcomes. In our case, the system presents four possible exits, and for low speeds the values of $S_{bb}$ largely exceed the $\ln 2$ threshold, as shown in Fig.~\ref{fig:diff_vx0}-(b). Note that the number of boxes at the boundary varies with the incident velocity (see Fig.~\ref{fig:diff_vx0}-(c)), this information should be taken into account for the experimental sampling. Moreover, we have checked that the $\ln 2$ criterion can be fulfilled for all the angles $\theta$ (except the limit cases $\theta=0^o,90^o$). If such values were obtained in experiments, it could be considered an experimental demonstration that the phase space is fractal.

Nevertheless, it is important to recall that the $\ln 2$ criterion detects fractals \textit{at a given resolution}. Indeed, given a finite resolution it is impossible to distinguish a real fractal from something which is not a fractal, but that looks like it at that resolution. The $\ln 2$ criterion presents a major advantage compared to other techniques like implementing directly the box-counting algorithm: it avoids the use of different scales of velocity and position, which, in the context of experiments with cold atoms, is fundamental. The $\ln 2$ criterion is a strong argument to test fractal structures using minimal requirements. Of course, we will detect fractal structures at the resolution that could be achieved with the experiments, which depends on the size of the wave packet compared to the size of the region of phase space considered. To illustrate this point we display in Fig.~\ref{fig:resolution} basins with different resolutions showing that experimentally achievable resolutions are enough to observe fractal basins.  

Some escape basins are not only fractal, but also posses the stronger property of Wada \cite{kennedy_basins_1991,daza_basin_2016}. This means, that all the basins have a common boundary separating them. The experimental evidence of the Wada property would be that in this regime every time that more than one branch is populated, all the branches are populated. If the experiment is in the Wada regime, we will never detect atoms escaping through only two or three different branches.

\section{\label{sec:Discussion}Discussion}

In this work, we have explained in detail how it is possible to use new techniques from nonlinear dynamics to analyze the chaotic dynamics of cold atoms directly from experiments. We have focused on a double guide configuration, where the atoms can escape through four different exits. In real experiments, we can measure the atom population in each branch, that is, the proportion of atoms that have escaped from the scattering region through each exit. Gathering this information through an appropriate Monte Carlo sampling, we can measure the basin entropy for a given set of parameters. This enables the characterization of the final state unpredictability associated to different experimental regimes. Using the same data set, for a suitable range of parameters, we have shown how the presence of fractal structures in phase space can then be detected. An interesting extension of the experimental setting is the inclusion of more guides. Indeed, with more exits, the $\ln 2$ criterion would be more easily fulfilled, and should facilitate the detection of fractality. Our study motivated by an experimental work therefore provides a new approach to investigate nonlinear effects such as fractal structures or even the intriguing Wada property with cold atom experiments.

In terms of applications, the tools developed here can be used more systematically to investigate the efficiency and robustness of the switch and splitter regimes of the crossed beam configuration in order to use it reliably as part of a matter wave circuit \cite{atomtronics}. The protocols that we propose have been designed for a direct implementation with state of the art experimental techniques. 

%This work lays the foundations for future research where experiments with cold atoms will be enriched by their dynamics characterization. Our study motivated by an experimental work provides a new approach to investigate nonlinear effects such as fractal structure or Wada property in a way perfectly accessible with state of the art experimental techniques. 

\section*{Acknowledgments}

This work was supported by Spanish Ministry of Economy and Competitiveness under Project No. FIS2013-40653-P and FIS2016-76883-P. Financial support from the Programme Investissements d'Avenir under the program ANR-11-IDEX-0002-02, reference ANR-10-LABX-0037-NEXT is also acknowledged.\\

\end{document}